\documentclass[runningheads]{llncs}
\pdfoutput=1
\usepackage{amssymb}
\usepackage{graphicx}
\usepackage{subfigure}
\usepackage{epstopdf}
\begin{document}
\title{NAIRS: A Neural Attentive\\ Interpretable Recommendation System \thanks{This paper was published as a demonstration paper on WSDM'19. In this version, we added a detailed related work section.}}
%
%\titlerunning{Abbreviated paper title}
% If the paper title is too long for the running head, you can set
% an abbreviated paper title here
%
\author{Shuai Yu\inst{1} \and
Yongbo Wang\inst{2} \and
Min Yang\inst{2} \and Baocheng Li\inst{2}\and  Qiang Qu\inst{2} \and Jialie Shen\inst{3}}
%

% First names are abbreviated in the running head.
% If there are more than two authors, 'et al.' is used.
%
\institute{School of Computer Science, Fudan University \and
Shenzhen Institutes of Advanced Technology, Chinese Academy of Sciences \and
School of Electronics, Electrical Engineering and Computer Science, Queen's University Belfast
}
\maketitle              % typeset the header of the contribution
\begin{abstract}
In this paper, we develop a neural attentive 
interpretable recommendation system, named NAIRS. A self-attention network, as a key component of the system, is designed to assign attention weights to interacted items of a user.  This attention mechanism can distinguish the importance of the various interacted items in contributing to a user profile. %, and it also provides interpretable recommendations.
Based on the user profiles obtained by the self-attention network, NAIRS offers personalized high-quality recommendation.  Moreover, it develops  visual cues to interpret recommendations. This demo application with the implementation of NAIRS enables users to interact with a recommendation system, and it persistently collects training data to improve the system. The demonstration and experimental results show the effectiveness of NAIRS.

\end{abstract}
\section{Introduction}
With the huge volumes of online information, attention has been continuously paid to recommender systems \cite{wang2017irgan,2017dynamic}. Item-based collaborative filtering (CF) is one of the most successful techniques in practice due to its simplicity, accuracy, and scalability~\cite{kabbur2013fism,SarwarKKR01,NingK11}. It profiles a user with the historically interacted items and recommends similar items in terms of user profiles.

Most of the existing item-based CF methods utilize statistical measures (e.g., cosine similarity) to estimate item similarities. %, and provide  recommendation by choosing items that are similar to the user's historical interacted items. 
However, the assumption of equal weights is often applied for the items in the measurement \cite{kabbur2013fism}. In other words, different items in the historical list are equally treated, which is not true for many of the real-world recommendation applications. 
On the other hand, interpretable recommendations are of increasing interest,  %in many application areas. 
which explain the underlying reasons for the potential user interest on the recommended items. 
Traditional methods often generate explanations from the textual data such as the content and reviews associated with the items~\cite{ZhangL0ZLM14,ChenW17,Bauman0T17,ChelliahS17}. Yet, 
generating reasons of recommendation remains unsolvable when the texts are unavailable.

Inspired by the recent successes of attention-based neural networks~\cite{bahdanau2014neural} in computer vision and natural language processing,  this paper proposes a neural attentive interpretable recommendation system (NAIRS) to alleviate the aforementioned limitations. 
The key to the design of NAIRS is an self-attention network that computes the attention weights of the historical items in a user profile according to their intent importance associated with the user's preferences. With the learned attention weights, NAIRS provides a high-quality personalized recommendation to users according to their historical preferences. Meanwhile, it interprets the  reasons of recommendation by visualizing the learned attention weights for the user's historical list. 
The function of personalized and interpretable recommendation assists users and manufacturer in identifying
results of interest and exploring alternative choices more efficiently.
In addition, NAIRS enables users to search for the users who have the similar results and search for the items which are similar to the chosen item. The two functions help the users to discover more potentially interesting items. 
Furthermore, NAIRS actively records users' interactive behaviors in the system, such as their input queries, liked items, and clicked results. 
\section{Related Work}
Recommender system is an active research field.
The authors of~\cite{bobadilla2013recommender,lu2015recommender}
described most of the existing techniques for recommender systems.
In this section, we briefly review the following major approaches that are  related to our work. 
\subsection{Item-based Collaborative Filtering}
Item-based collaborative filtering \cite{sarwar2001item} is one of the most successful techniques in the practice of recommendation due to its simplicity and attractive accuracy. 
The main idea behind item-based CF is that the prediction of a user
$u$ on a target item $i$ depends on the similarity of $i$ to all
items the user $u$ has interacted with in the past. 
In \cite{ning2011slim}, the authors proposed a method named SLIM (short
for Sparse Linear Method), which learned item similarities by
optimizing a recommendation-aware objective function. SLIM minimized the loss between the original user-item
interaction matrix and the reconstructed one from the
item-based CF model. 
FISM \cite{kabbur2013fism} was one of the most widely used collaborative filtering method, which achieved the state-of-the-art performance among the item-based methods. In its standard setting, the prediction of a user $u$ to an item $i$ is calculated by the inner product of the historical items and the target item.   \cite{wang2018attention} proposed an attention-based transaction embedding (ATEM) model.  It was a shallow wide-in-wide-out neural network, which learned an attentive context embedding that is expected to be most relevant to the next choice over all the observed items in a transaction.  \cite{he2018nais}  leveraged historical items as attention source to calculate the relationship between the historical items and new-coming item.  \cite{DBLP:conf/ijcai/ZhangYSWLD18} proposed an integrated network to combine non-linear transformation with latent factors. 
\subsection{Deep Learning for Recommender Systems}
These traditional MF methods for recommender systems are based on
the assumption that the user interests and movie attributes are near static,
which is however not consistent with reality. \cite{koren2010collaborative} 
discussed the effect of temporal dynamics in recommender systems
and proposed a temporal extension of the SVD++ (called TimeSVD++)
to explicitly model the temporal bias in data. However, the features
used in TimeSVD++ were hand-crafted and computationally expensive to obtain. Recently,
there have been increasing interests in employing recurrent neural
network to model temporal dynamic in recommender systems. For example, \cite{hidasi2015session}  applied recurrent
neural network (i.e. GRU) to session-based recommender systems. This work
treated the first item a user clicked as the initial input of GRU.
Each follow-up click of the user would then trigger a recommendation
depending on all of the previous clicks. \cite{wu2016recurrent} 
proposed a recurrent neural network to perform the time heterogeneous
feedback recommendation. \cite{wu2017recurrent}  used LSTM autoregressive model for the user and movie dynamics and employed matrix factorization to model
the stationary components that encode fixed properties. 
%Different from their work,  we use GAN framework to leverage the MF and RNN approaches for top-$n$ recommendation,  aiming to generate plausible and high-quality recommendation lists.  

To address the cold start problem in recommendation, \cite{cui2016visual}  presented a visual and textural
recurrent neural network (VT-RNN), which simultaneously learned the sequential latent vectors of users' interest
and captured the content-based representations that contributed to address the cold-start issues.

\subsection{Attention Network For Recommendation}
Attention mechanisms have recently become an essential part in deep neural networks, which equip a deep neural network with the ability to focus on a subset of its inputs (or features).  
Self-Attention Network \cite{vaswani2017attention} is a special case of the attention mechanism, which uses a token embedding from the source input itself as the attention source to calculate the distributed representation of the input sequence.  
It relates elements at different
positions from a single input sequence by computing the attention between each pair of tokens. 
Expressive performance have been achieved by self-attention mechanism for modeling both long-range and local dependencies of the input sequence. 
%Moreover, it has much faster computation speed and fewer parameters than RNN. 
Recently, remarkable success has been achieved by self-attention in a variety of  tasks,
such as reading comprehension \cite{hu2017reinforced}
and neural machine translation \cite{vaswani2017attention}.
ATRank \cite{zhou2018atrank} can model with heterogeneous
user behaviors using only the attention model. Behaviors interactions are captured using self-attention in multiple semantic spaces. 
%The model can also perform multi-task that predict all types of user actions using one unified model, which shows comparable performance with the highly optimized individual models.
User preferences often evolve over time, thus modeling their temporal dynamics is essential for recommendation. 
\cite{pei2017interacting} proposed
the Interacting Attention-gated Recurrent Network (IARN)\cite{pei2017interacting} to accommodate temporal context for better recommendation. IARN can
not only accurately measure the relevance of individual time steps
of user and item history for recommendation, but also capture the dependencies between user and item dynamics in shaping user-item interactions

The studies most similar to ours are proposed in \cite{pei2017interacting,ying2018sequential,zhou2018atrank,hetkde}.  \cite{ying2018sequential} proposed a novel two-layer hierarchical attention network (SHAN) to recommend the next item the user might be interested in. Specifically, the first attention layer learns user long-term preferences based on the representations of historical purchased items, and the second layer outputs final user representation through coupling user long-term and short-term preferences.
\cite{zhou2018atrank} proposed
an attention based user behavior modeling framework (ATRank). Heterogeneous user behaviors are considered in the model
that project all types of behaviors into multiple latent semantic spaces, where influence can be made among the behaviors via self-attention.
\cite{hetkde} employed an item-side interactive neural attention network (NAIS), which assigned different weights on historical items.
Our model differs from NAIS in several aspects. First, we employ a self-attention mechanism to learn the representations of users, instead of calculating the attention scores with respect to specific items, as in \cite{hetkde}. 
Second, in practice, we can recommend items based on the pre-computed representations of users in real-time, while\cite{hetkde} needs to calculate attention weights every time.
In addition, NAIRS can provide user profiles that play a crucial role in broad applications. 
More importantly, this demonstration paper provides the live demonstration and prototype of the interpretable recommendation system.

\section{Core Algorithm}
We denote a user-item interaction
matrix as $\mathbf{R} \in \mathbb{R}^{M\times N}$, where M and N are the number of users
and items, respectively.
We use $\mathcal{R} =\{(i,j)|\mathbf{R}_{ij} = 1\}$ to denote the set of user-item
pairs and  use $\mathcal{R}_{u}^{+}$ to denote the set of items that user $u$ has interacted with.
As described in \cite{kabbur2013fism}, each item has two embedding vectors \textbf{p} and \textbf{q}
to distinguish its role of history item and prediction target.
\begin{figure}
	\centering
	\includegraphics[scale=0.4]{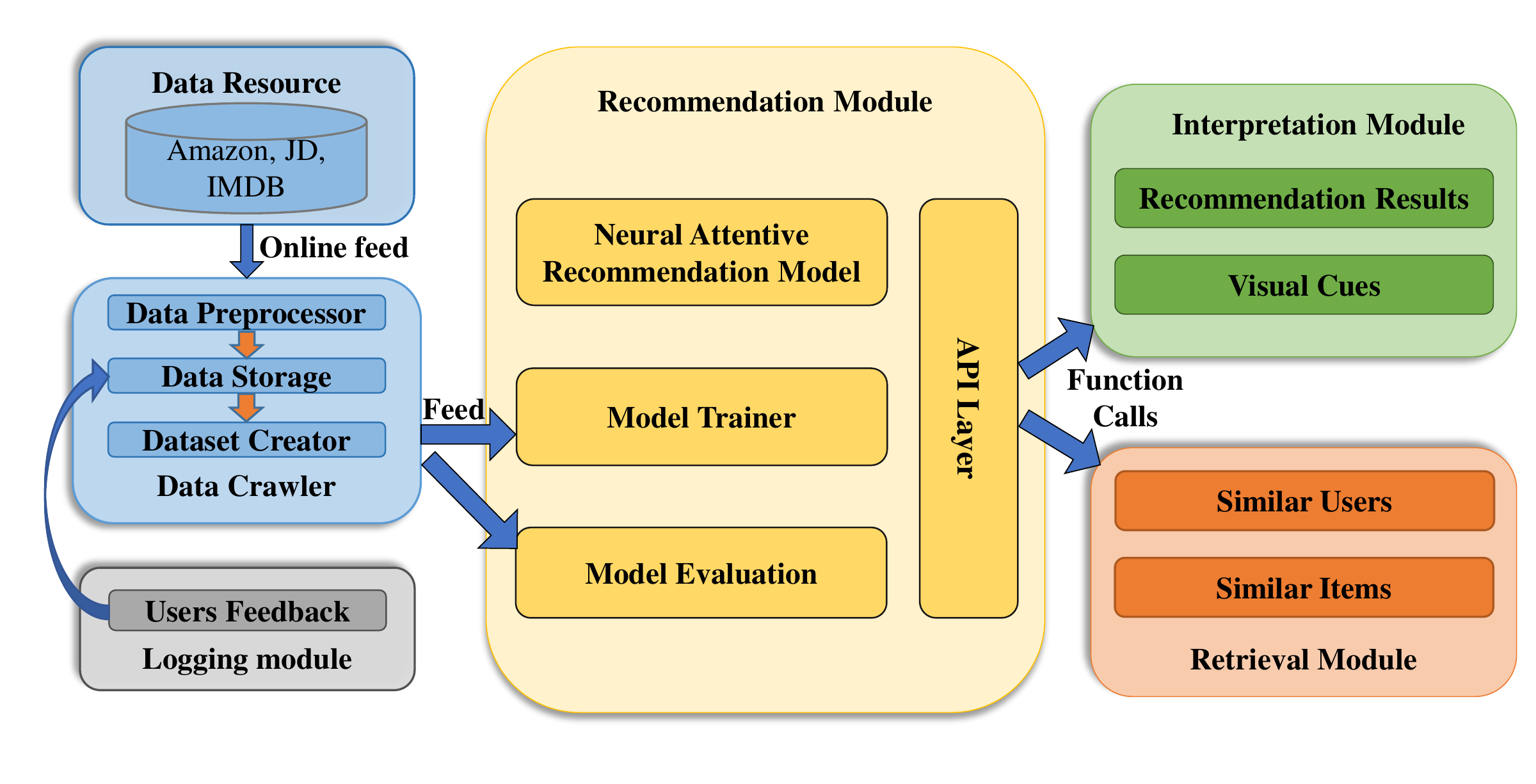}
	\caption{Architecture overview of NAIRS.}\label{fig.overview}
\end{figure}
%\yangmin{modify later}
The FISM \cite{kabbur2013fism} is  one of the most widely used collaborative filtering method, which achieves the state-of-the-art performance among the item-based methods. In its standard setting, the prediction
of a user $u$ to an item $i$ can be calculated as below:
\begin{equation}\label{fismeq}
\hat{r}_{ui}=b_u+b_i+ (\frac{1}{|\mathcal{R}_u^+|^{\alpha}}\sum_{j\in \mathcal{R}_u^+}\mathbf{p}_j^{T}) \cdot\mathbf{q}_i, 
\end{equation}
where $b_u$ and $b_i$ denote the user and item biases, respectively.

Despite the effectiveness of FISM, 
we argue that its performance is hindered by assigning equal weight to each interacted item.
To address this limitation, we propose a neural attentive network to assign different weights to the items according to their intent importance. Mathematically, the prediction of user $u$ to item $i$ can be calculated as 
\begin{equation}\label{yui}
\hat{r}_{ui}=b_u+b_i+(\sum_{j\in \mathcal{R}_u^+}\alpha_{uj}\mathbf{p}_j^{T})\cdot\mathbf{q}_i,
\end{equation}
where $\alpha_{uj}$ is the attention weight of item $j$ in contributing to user $u$'s representation.
Specifically, we exploit self-attention to learn the representation of user $u$, each of the historical items learns to align to each other. The weight $\alpha_{uj}$ of each historical item j is computed by
\begin{equation}\label{attention}
	\alpha_{uj}=\frac{exp(e(\mathbf{p}_j))}{[\sum_{k \in \mathcal{R}_u^+}exp(e(\mathbf{p}_k))]^{\beta}}
,
\end{equation}
\begin{equation}
    	e(\mathbf{p}_j)=\mathbf{V}^{T}g(\mathbf{W}\cdot\mathbf{p}_{j}+b)
\end{equation}
where $e(\mathbf{p}_j)$ is an alignment model which scores the contribution of item $j$ to the representation of user $u$. To form a proper probability distribution over the items, we normalize the scores across the items using \emph{softmax} function and get attention score $\alpha_{uj}$.  $\beta$ is a smoothing hyper parameter that will be discussed later in this section. $V$ and $W$ are the weight matrices, and $g$($\cdot$) is the activation function. 

In practice, the standard attention network fails to learn from users' historical data and perform accurate recommendation. By analyzing the attention weights outputted by the model, we reveal that the performance of the model
is largely hindered by the \emph{softmax} function, due to the large variances on the
lengths of user histories. The attention weights of the items from long history list are largely decreased.
To address this problem, we introduce a new symbol $\beta$ to smooth the denominator of the original attention formula\footnote{This smoothing method is similar to that in \cite{hetkde}. However, we did this work independently and this work (which was initially submitted to SIGIR-18) was done before the publication of \cite{hetkde}.}. $\beta$ can be set in a range of $[0, 1]$. If $\beta=1$, then Eq. (3) degenerates into the original \emph{softmax}. One typically chooses the value of $\beta$ between zero and one. This smooth setting leads to much better
performance than standard \emph{softmax} function.

Following the strategy in the previous work \cite{he2017neural}, we treat the observations as positive instances and randomly sample the unobserved items as negative instances.
Cross entropy is adopted as the objective function, which minimizes the regularized log loss:
\begin{equation}\label{loss}
L=-\frac{1}{N}\left[\sum_{(u,i)\in \mathcal{R}^+_u}r_{ui}\log\sigma(\hat{r}_{ui}) + 
\sum_{(u,i)\in \mathcal{R}^-_u} (1-r_{ui})\log(1-\sigma(\hat{r}_{ui}))\right] \\
+\lambda\left\Vert\theta\right\Vert^2 
\end{equation}
where $N$ denotes the number of the training instances, $\theta$ denotes the parameters of the model.
\section{System Architecture}
The proposed NAIRS, overviewed in Figure~\ref{fig.overview}, consists of five main modules.
(1) The \emph{Data crawler} module collects  
user interactive information from various websites such as Amazon, Jindong, and IMDB. 
(2)  The \emph{Recommendation} module produces recommendation results and interpretable partial scores of user-item pairs. 
(3) The \emph{Interpretation} module visualizes the interpretable reasons of the recommendation by scoring user's historical list. 
(4) The \emph{Retrieval} module enables the users to i) find people with similar preferences (with historical lists) and ii) explore the items that are similar to a user-specified item. This module effectively assists the users in finding more items in which they may be interested. 
(5) The \emph{Logging} module collects user behaviors from the system, such as chosen items. The logging information is utilized to further improve the recommender system. 
In the rest of this section, we elaborate each module of the above.
\begin{figure*}
	\centering
	\includegraphics[scale=0.32]{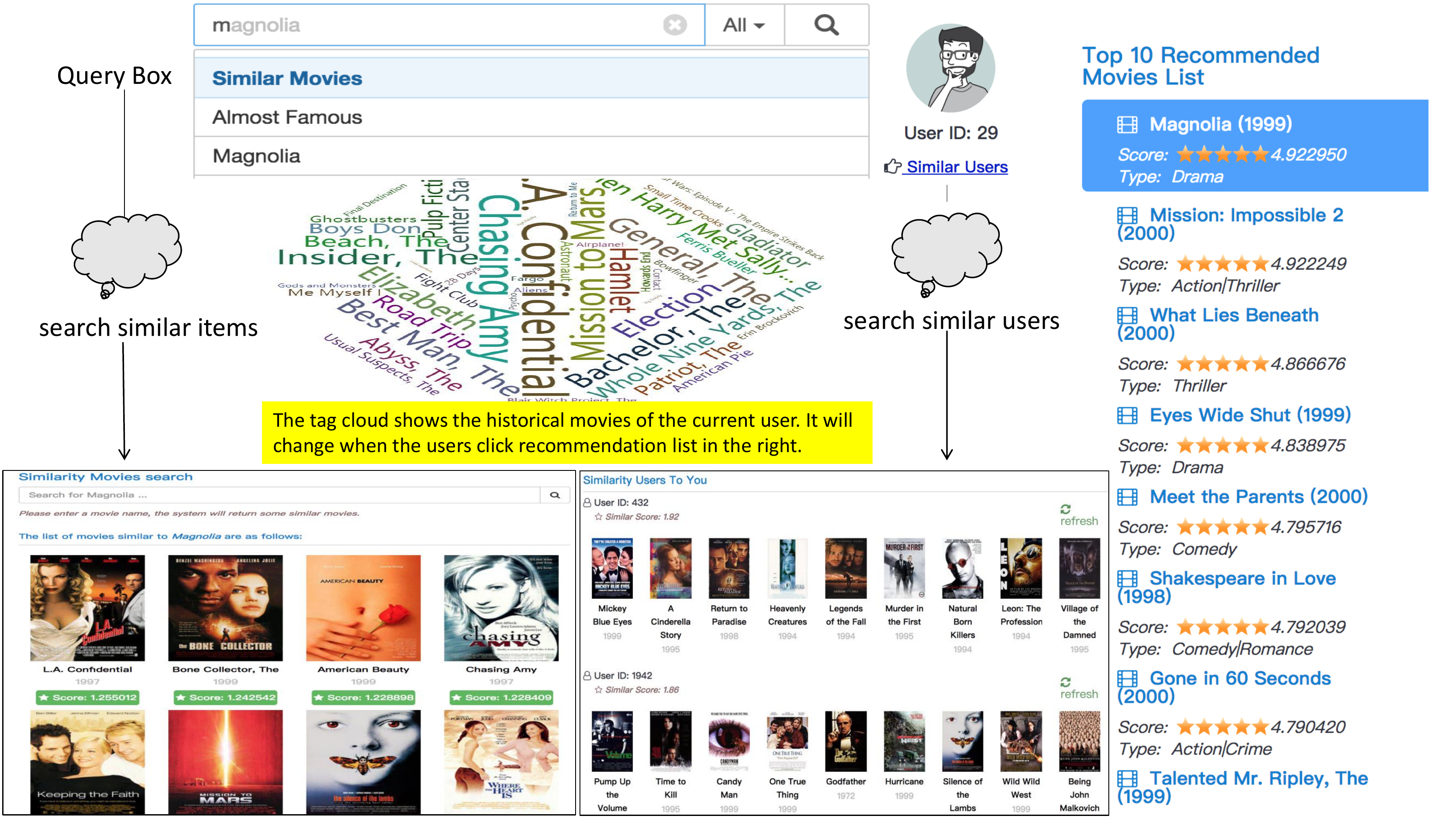}
	\caption{The \emph{Neural Attentive Interpretable Recommendation System}. The top part shows the \emph{Interpretable Recommendation} module, the bottom part shows the \emph{Retrieval Module}. The user's historical interacted movies are displayed in the tag cloud. When the user clicks a movie in the recommendation list, the related movies in the tag cloud will become bigger. The user can either search similar items in the query box or click the movie's name in the tag cloud. The user can also click the link bellow the user logo to explore the users who have similar interests to her/him.}\label{figure2}
\end{figure*}
\subsection{Data Crawler Module}
The data crawler collects three types of user-item interactive data: (i) movie rating data from IMDB\footnote{https://www.imdb.com}; (2) books rating data from Amazon\footnote{https://www.amazon.com}; and (3) daily goods rating data from Jingdong\footnote{https://www.jd.com}. For movie rating data, users are selected at random for inclusion. All selected users have rated at least one movie. For book rating data, we focus on the top 1000 popular books. We also collect six categories of daily goods including clothes, shoes, cosmetics, foods, toys, and smart phones. In this work, all user-sensitive information is removed.

\subsection{Recommendation Module}
We implement the interpretable recommendation algorithm introduced in Section 2 to perform top-$n$ item recommendation.
Our recommendation model is implemented with the TensorFlow\footnote{https://www.tensorflow.org/} library and trained on a NVIDIA Titan Xp GPU.  After training, we can obtain the user and item representations for each user and item, which are then used to predict the rating scores and assign weights to items in user's historical list for interpretation.
In addition, we can obtain the \emph{similar users} and \emph{similar items} results easily with the learned user and item representations.
The results learned by Recommendation module can be directly used by the Interpretation module and the Retrieval module. 

Note that during the bootstrap process NAIRS provides users a navigation page in which the users can choose the items that they are interested in. This process can alleviate the cold start problem in recommendation to some extent, especially for new users. 
\subsection{Interpretation Module}
Given a user $u$, the historical items $\mathcal{R}_{u}$, and a recommended item $q_i$, the \emph{Interpretation} module  provides the top-$n$ recommendation results and interprets the reasons of the recommendation by visualizing the attention scores of user $u$'s historical list $\mathcal{R}_{u}$. In particular, we support the users to add interested movies into their profile list or delete the movies they do not like.  NAIRS then demonstrates the recommendation results (on the right of the interface) and interprets the reason of each recommendation with a tag cloud (in the center of the interface), as shown in Figure~\ref{figure2}. The importance of each item in the user's historical list is shown with various font sizes. The larger the item names, the more important the items in contributing to the recommendation. For example, the movie \emph{Men in Black} is recommended based on \emph{Nikata} in the user's historical list which has the highest attention score. We show that these two movies  both belong to the action movie category. On the other hand, the movie \emph{Escape from New York} contributes little to recommend movie \emph{In the Army Now} since they belong to different categories.
\subsection{Retrieval Module}
\subsubsection{Similar Users}
The \emph{Similar Users} module can assist  end users to find  other users who have similar interests. 
This module plays an important role in helping the users who might not know exactly what they are looking for to discover potentially interesting items based on the observation that people who agree in the past are likely to agree again. 
In order to overcome the insensitive of average value, we calculate the similarity between users with \emph{adjusted cosine similarity} as follows:
\begin{equation}
sim(u,u')=1+\frac{\sum_k{(\mathbf{u}_{k}-\mathbf{\bar u}) \mathbf{\cdot} (\mathbf{u_{k}^{'}}} - \mathbf{\bar u_k})}{\sqrt{\sum {(\mathbf{u}_{k}-\mathbf{\bar u}})^{2}} \sqrt{\sum {(\mathbf{u}^{'}_{k}-\mathbf{\bar u'}})^{2}}}, 
\end{equation}
where $\bar{u}$ and $\bar{u'}$ are the average values over the user's embedding dimensions. Note that we map the value space of the similarity from [-1, 1] to [0, 2] to provide positive similarity scores for better visualization. 
As shown in Figure~\ref{figure2}, we visualize the similar users with their historical lists.
In addition, we provide the ``like'' and ``dislike'' buttons for users to  select/filter the displayed items. 
These feedback information can be used to update our Recommendation module. In NAIRS, the calculated similarities are cached after updating the model to speed up the results retrieval process.

\subsubsection{Similar Items}
%Similar to the  \emph{Similar Users} component, 
Intuitively, a user is likely to have similar level of interest for similar items. The \emph{Similar items} module finds items similar to the items liked or chosen by the user. In particular, we provide the end user with a search window for searching any items in the system. Then the items whose similarities are above a threshold are returned as the search results. 
The similarity between items is calculated as follows:
\begin{equation}
sim(i,i')=1+\frac{\sum_k{(\mathbf{i}_{k}-\mathbf{\bar i}) \mathbf{\cdot} (\mathbf{i_{k}^{'}}} - \mathbf{\bar i_k})}{\sqrt{\sum {(\mathbf{i}_{k}-\mathbf{\bar i}})^{2}} \sqrt{\sum {(\mathbf{i}^{'}_{k}-\mathbf{\bar i'}})^{2}}}.
\end{equation}
Similar to the \emph{Similar Users} component, the similarities between items are cached in the system. When the end user requests similar items, we can obtain the results in $O(1)$ time.
\section{DEMOSTRATION}
\subsection{Demonstration Setup}
The \emph{NAIRS} prototype has client and server ends. Clients can access the system by web, mainly for rendering recommendation, interpretation, search, and query results. The server is deployed on Apache Tomcat, which performs the recommendation algorithm and communicates with clients.
\subsection{Walkthrough Example}
The NAIRS demo consists of the following steps:

\textbf{Step 1:} The user can access to the system by web either on PCs or smart phones. After logging onto the system, the user can select a kind of recommendation service from three categories: movies, books, and daily goods.

\textbf{Step 2:} If the user is new to the system, a collection of randomly chosen items are presented, and the user is asked to choose some items in which the user is interested. After submitting the chosen items, the system offers the top-10 recommendation lists based the chosen items. Furthermore, the system shows tag cloud of the user profile, which reveals why the system recommends the specific items to the user. The user can click any item in the recommendation list, and the user profile tag cloud change accordingly.

\textbf{Step 3:} The user can query other users that have similar interests by clicking the ``similar users'' button. 
Then the similar users with their historical lists are returned to the user, and the user can choose to follow them and find the potentially interesting items via this function. 

\textbf{Step 4:} The user can also search the items similar to the item inputed by the user. If our system has items for which the user search, similar items are returned; otherwise, a warning message is shown.
Note that to enhance user experience, we implement an Auto-suggestion query box.\section{Quantitative Evaluation}
In this section, we evaluate the performance of NAIRS quantitatively, then we investigate
the interpretation of the proposed system.
\begin{figure}
	\centering
	
	\subfigure[Movielens HR]{
		\begin{minipage}{4cm}\centering
			\includegraphics[scale=0.2]{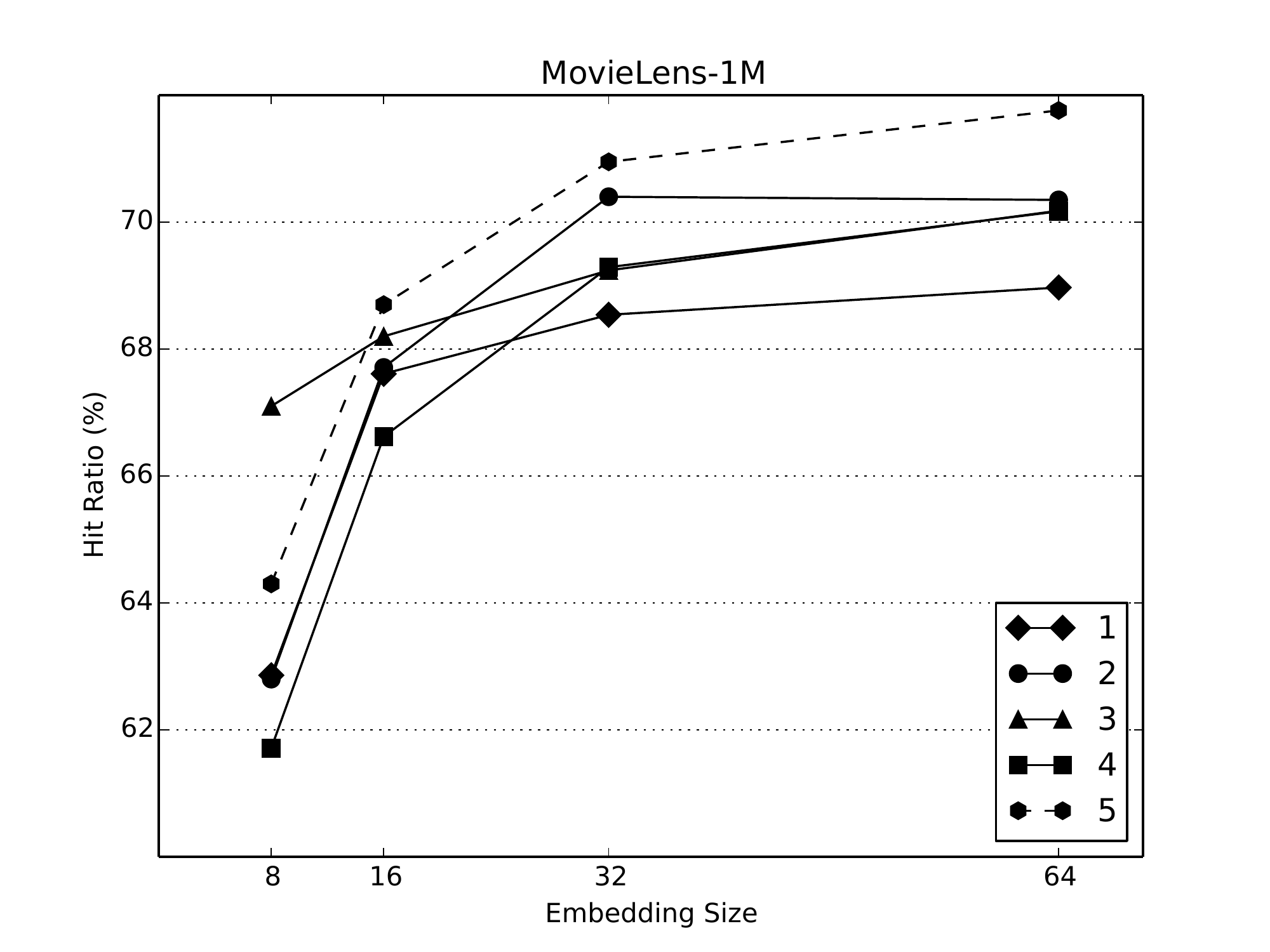}
	\end{minipage}}
	\subfigure[Movielens NDCG]{
		\begin{minipage}{4cm}\centering
			\includegraphics[scale=0.2]{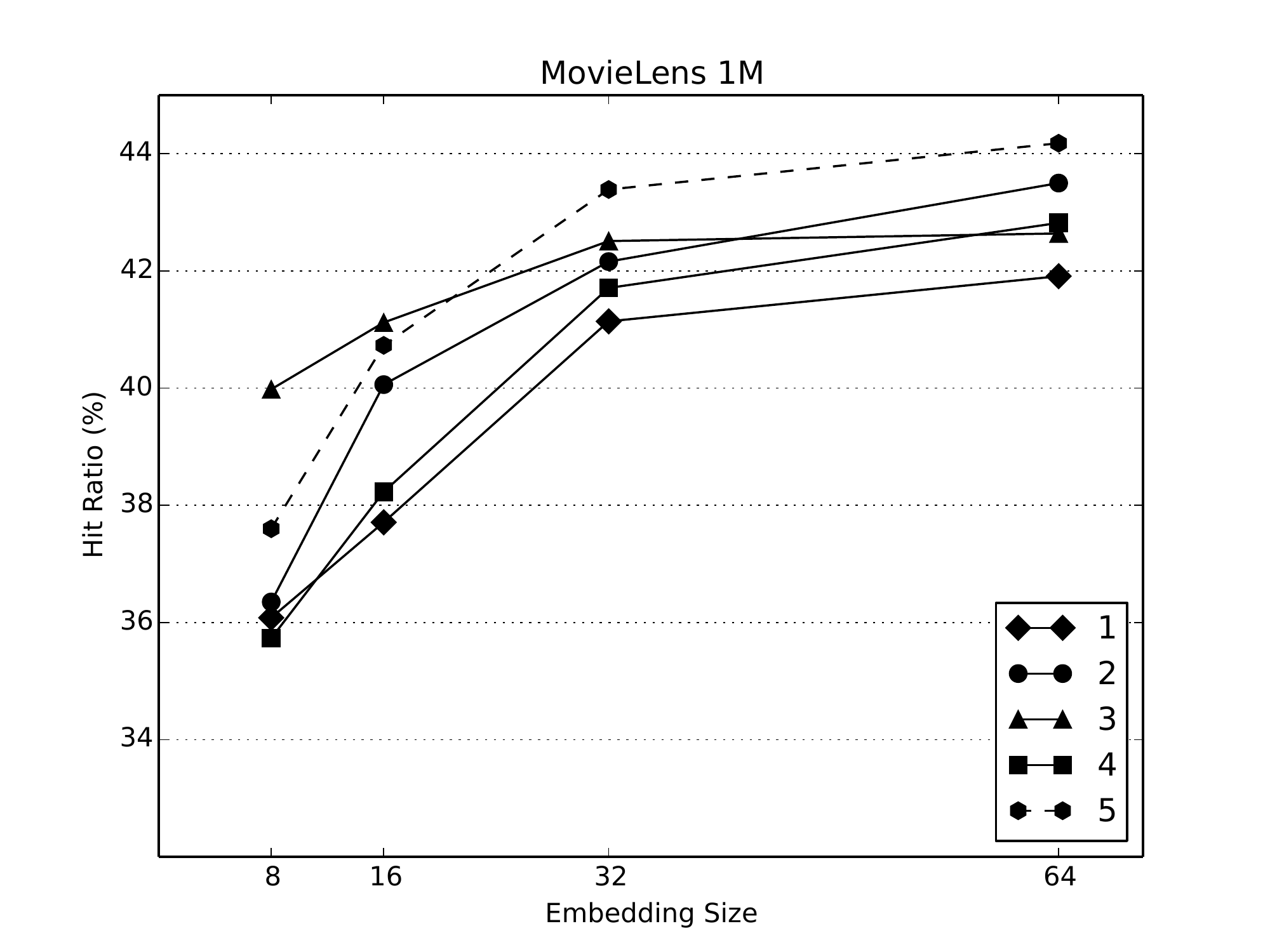}
	\end{minipage}}
	\subfigure[Pinterest HR]{
		\begin{minipage}{4cm}\centering
			\includegraphics[scale=0.2]{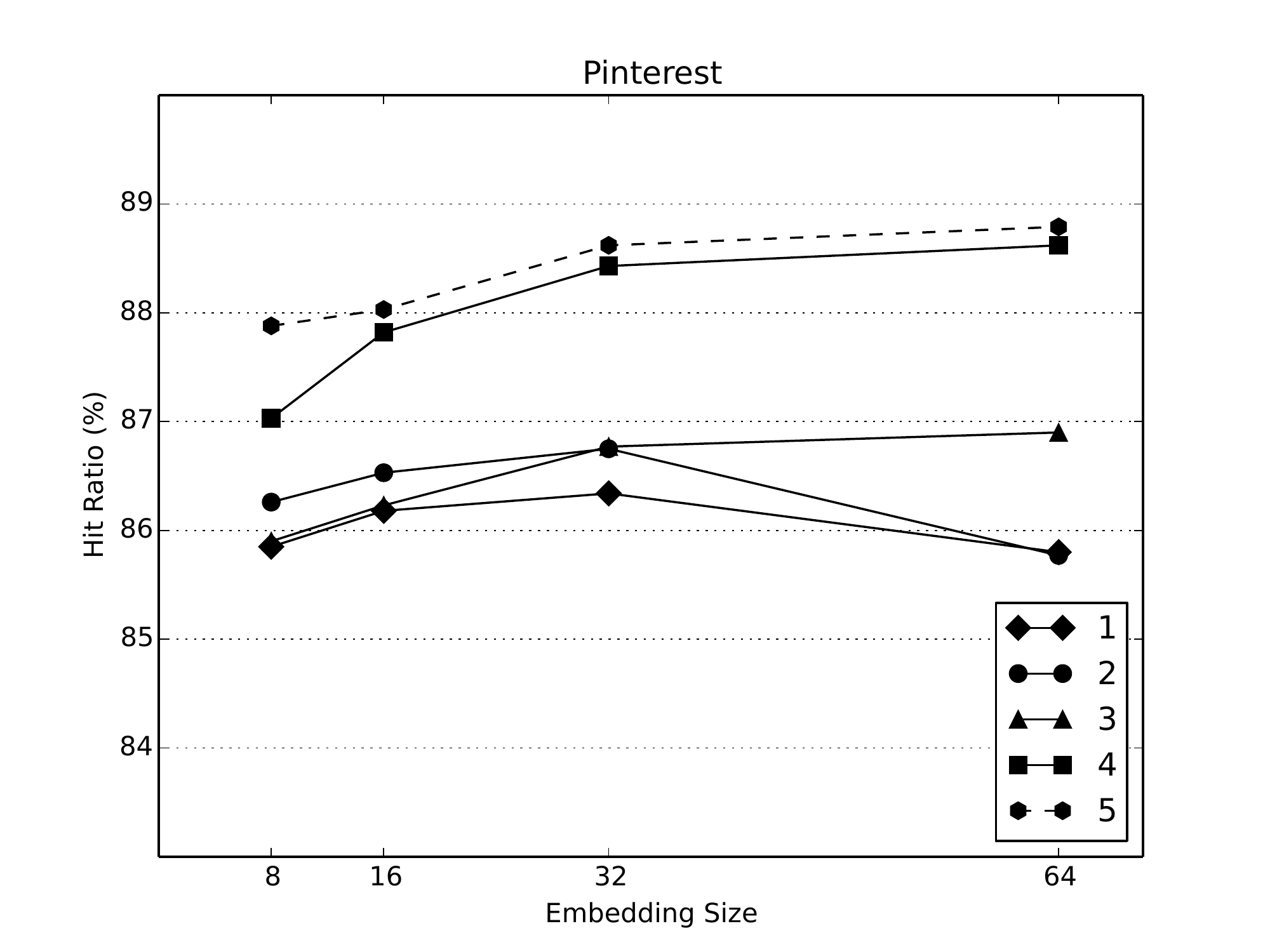}
	\end{minipage}}
	\subfigure[Pinterest NDCG]{
		\begin{minipage}{4cm}\centering
			\includegraphics[scale=0.2]{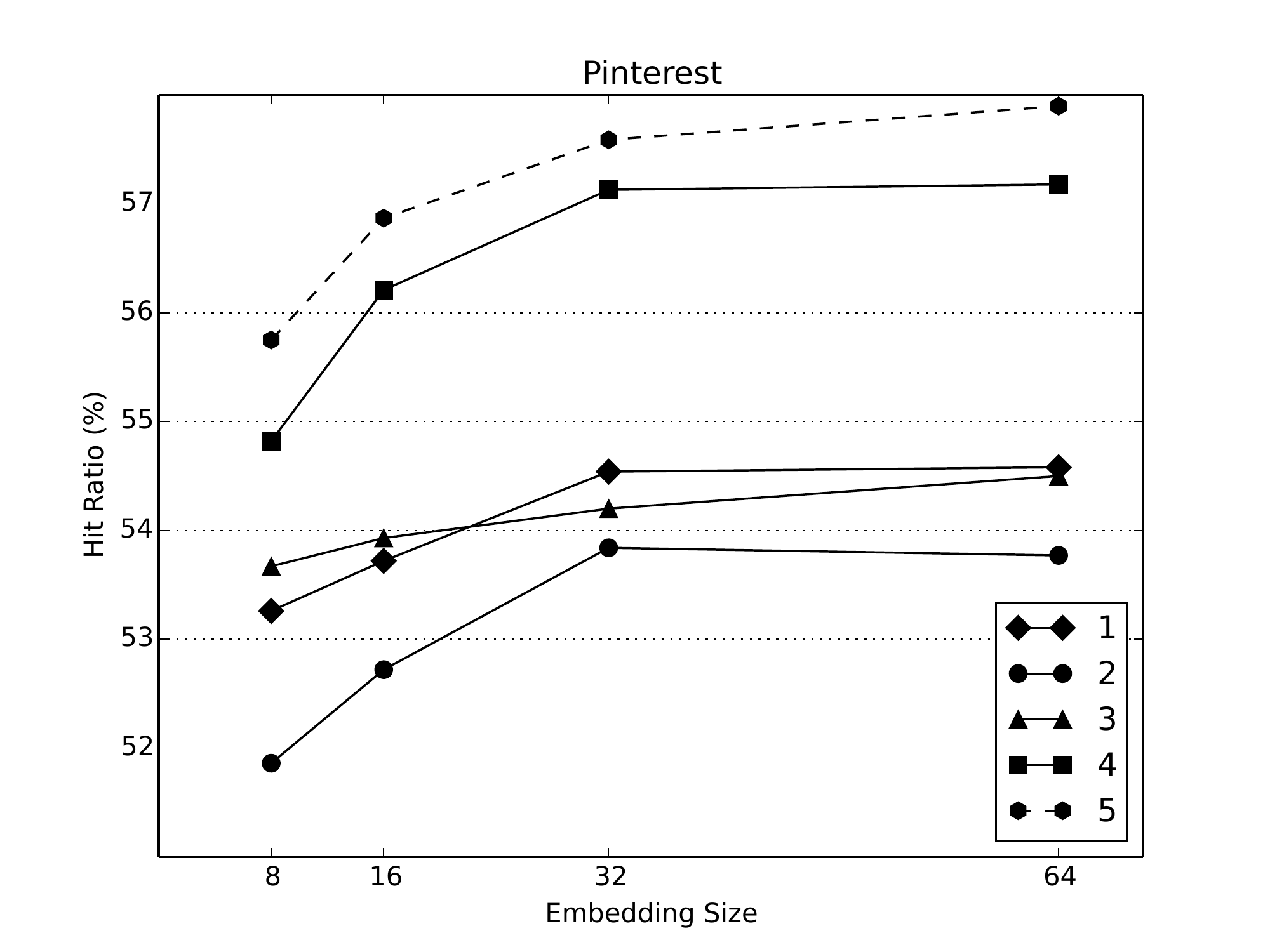}
	\end{minipage}}
	%\vspace{-10px}
	\caption{Performance comparison.}
	\label{figure4}
	\vspace{-0.3cm}
\end{figure}
We conduct experiments on two widely used datasets: Movielens-1M and Pinterest, as the ones
used in the study~\cite{he2017neural}. 
The results are judged with hit ratio(HR) and Normalized Discounted Cumulative Gain(NDCG), which have been widely used in top-$n$ recommendation~\cite{kabbur2013fism,he2017neural}. 
NAIR is compared with several baseline methods including
MF-BPR~\cite{rendle2009bpr}, MF-eALS~\cite{he2016fast},
FISM~\cite{kabbur2013fism}, and MLP~\cite{he2017neural}.

The experimental results are shown in Figure \ref{figure4}. We observe that our method outperforms other competitive methods for both of the datasets, which shows the effectiveness of the proposed approach on top-$n$ recommendation used in the demonstration.

%
% ---- Bibliography ----
%
% BibTeX users should specify bibliography style 'splncs04'.
% References will then be sorted and formatted in the correct style.
%
\bibliographystyle{splncs04}
\bibliography{mybib}

\end{document}